\documentclass[12pt]{article}
\usepackage{amsmath}
\usepackage{amssymb}
\usepackage{mathrsfs}
\def\tuc {}

\def\ba{\begin{eqnarray}}
\def\ea{\end{eqnarray}}
\def\be{\begin{equation}}
\def\ee{\end{equation}}

\def\eqn{equation}

\def\tfn{transformation}

\def\sm{sigma model}
\def\pl{Poisson--Lie}

\def\dd{Drinfeld double}

\def\3dial{three-dimensional}
\def\-1{^{-1}}
\def\half{\frac{1}{2}}

\def\real{\mathbb{R}}

\def \unit {{\bf 1}}

\def\e{{\rm e}}

\def\cg{{\mathfrak g}}

\def\wt{\tilde}
\def\sm{sigma model}

\def\pltd{Poisson--Lie T-dualit}

\begin{document}
\title{\bf{New solvable  sigma models in plane--parallel wave background}}
\author{{Ladislav Hlavat\'y\footnote{hlavaty@fjfi.cvut.cz}, Ivo Petr\footnote{ivo.petr@fjfi.cvut.cz}}}\maketitle
{Czech Technical University in Prague, Faculty of Nuclear Sciences
and Physical Engineering,  B\v rehov\'a 7, 115 19 Prague 1, Czech
Republic}
\begin{abstract}{We explicitly solve the classical equations of motion for strings
in backgrounds obtained as non-abelian T-duals of a homogeneous isotropic plane--parallel
wave. {To construct the dual backgrounds, semi-abelian Drinfeld doubles are used which
contain the isometry group of the homogeneous plane wave metric. The dual solutions are
then found by the Poisson--Lie transformation of the explicit solution of the original
homogeneous plane  wave background. Investigating their Killing vectors, we have found that the dual backgrounds can be transformed to the form of more general
plane--parallel waves.}

}
\end{abstract}
\tableofcontents
\section{Introduction}
Sigma models satisfying supplementary conditions can serve as models
of string theory in curved and
time-dependent backgrounds. {Finding the solution} of sigma models in 
such backgrounds is often very complicated, not to say impossible. That is why every
solvable case attracts considerable attention. An example of such a solvable model is a
string theory in homogeneous plane--parallel wave background investigated in Ref.
\cite{papa}. {The general relativistic statement that any spacetime has a plane wave as a
limit \cite{Penrose} was reformulated for string theories in Ref. \cite{Gueven}. The
authors of Ref.  \cite{papa} mention the relevance of the homogeneous plane wave
backgrounds as Penrose limits of cosmological and p-brane backgrounds, and  study the
possibility of string propagation through the singularity of the metric}. The classical
string is solved there in terms of Bessel functions and subsequently quantized.

{Ref. \cite{Gueven} also discusses the relation between Penrose limits and T-duality. Since
the homogeneous plane--parallel  wave metric investigated in Ref. \cite{papa} has rather
large group of isometries, dual backgrounds can be constructed via non-abelian T-duality
introduced in Ref. \cite{delaossa:1992vc} and investigated further in the context of WZW
models or gravitational instantons in Refs.
\cite{AlvaAlvaGaBarLoz},\cite{AlvaAlvaGaumeLozano}. Studied originally for the bosonic
strings, the non-abelian T-duality
 was recently extended to superstring backgrounds including the Ramond fluxes
\cite{1012.1320 sfethomp},
\cite{1104.5196 LCST}, with the supersymmetric  plane--parallel wave background studied in Ref. \cite{1205.2274 ILCS}
. However, other applications  of non-abelian duality in
cosmology can be found, see e.g. Ref. \cite{0210211 BosMoham}. }

The non-abelian T-duality provides us with a prescription how to relate solutions of sigma
models with apparently different target space geometries. Knowledge of the solution of the
original model gives us in principle the possibility to solve the equations of motion of
strings in dual backgrounds. In the case of  the homogeneous plane wave we were able to
find the explicit solutions. In the following, we shall understand the non-abelian
T-duality as a special case of Poisson--Lie T-duality \cite{klise}, which solved the
problem that the non-abelian dual may have less symmetries than the original model, making
it impossible to find reverse non-abelian T-duality transformation by the procedure
of Ref. \cite{delaossa:1992vc}.

The backgrounds resulting from the {Poisson--Lie} T-duality are expressed in the so
called group coordinates that may hide their commonly used forms. To find these, {one can
analyze the the Killing vectors, which} provide important information about the symmetries
of the dual backgrounds, and meanwhile suggest a way to find the coordinate \tfn s for
transition to the commonly used forms {of the dual metrics}.

{The plan of the paper is the following. In the next section we review the method of the
\pltd  y that is used later as a tool for construction of the dual models and their
solution. Relevant results concerning homogeneous isotropic plane wave and the solution of
classical string equations of motion obtained in \cite{papa} are summarized in the third
section. Section 4 describes the two particular Drinfeld doubles which are dealt with later
in sections 5 and 6, where the corresponding dual backgrounds are solved. In their
subsections the symmetries of the backgrounds are studied, and the
plane--parallel wave form of the background is revealed through an appropriate coordinate
transformation}.

\section{Elements of Poisson--Lie T-dual sigma models}\label{secPLT}
 The basic concept used for construction
of {mutually} dual sigma models is a \dd {} -- a Lie group with an additional structure.
{More precisely, there are two equally dimensional subgroups $G, \tilde{G}$ in the
\dd{} $D$, such that  Lie subalgebras ${\mathfrak{g}}$, ${\tilde{\mathfrak{g}}}$ are
isotropic subspaces of the Lie algebra ${\mathfrak{d}}$ of the \dd. The Drinfeld double
suitable for a given sigma model living in curved background can sometimes be found from
the knowledge of the symmetry group of the metric. In the case that} the metric has
sufficient number of independent Killing vectors, the subgroups of the isometry group can
be taken as one of the subgroups of the \dd. The other one then {has to} be chosen
Abelian, in order to satisfy the conditions of dualizability. {We shall focus on the
case that the isometry subgroup acts freely and transitively on the manifold, which is the
case usually referred to as the atomic duality \cite{klim:proc}}. Let us summarize the
details.

Let $G$ be a Lie group and ${\mathfrak{g}}$ its Lie algebra. Sigma model on the group $G$
is given by the classical action
    \begin{align}
    \label{sma}
        &S_{F}[\phi]=
-\int d\sigma_+d\sigma_-\,(\partial_{-}\phi^{\mu}F_{\mu\nu}(\phi)\partial_{+}\phi^{\nu})=\\
=\nonumber \half \int d\tau
d\sigma\,&\left[-\partial_{\tau}\phi^{\mu}G_{\mu\nu}(\phi)\partial_{\tau}\phi^{\nu}+
\partial_{\sigma}\phi^{\mu}G_{\mu\nu}(\phi)\partial_{\sigma}\phi^{\nu}
-2\partial_{\tau}\phi^{\mu}B_{\mu\nu}(\phi)\partial_{\sigma}\phi^{\nu}\right],
    \end{align}
where  $F$ is a second order tensor field on the Lie group $G$, {with the metric and the
torsion potential}
    \[ G_{\mu\nu}=\half(F_{\mu\nu}+F_{\nu\mu}),\        B_{\mu\nu}=\half(F_{\mu\nu}-F_{\nu\mu}),
    \]
    {and the worldsheet coordinates}
    \[
   \sigma_+=\frac{1}{\sqrt{2}}(\tau+\sigma),\ \sigma_-  =\frac{1}{\sqrt{2}}(\tau-\sigma).
    \]
    The functions
$\phi^{\mu}$ 
are determined by the composition $\phi^{\mu}(\tau,\sigma)=x^{\mu}( g(\tau,\sigma))$, where
$g:{\mathbb{R}^{2}}\ni(\tau,\sigma)\mapsto g(\tau,\sigma)\in G$ and
$x^{\mu}:\mathbf{U}_{g}\rightarrow \mathbb{R}$ are components of a coordinate map of
neighborhood $\mathbf{U}_{g}$ of element $g(\tau,\sigma)\in G$.

The condition of dualizability of sigma models on the level of  the Lagrangian is given by
the condition \cite{klise} {for the Lie derivative of $F_{\mu\nu}$}
    \be\label{dc}
        {\mathcal{L}}_{v_{i}}F_{\mu\nu}=F_{\mu\kappa}v_{j}^{\kappa}\tilde{c}^{jk}_{i}v^{\lambda}_{k}F_{\lambda\nu},
    \ee
where $\tilde{c}_{i}^{jk}$ are structure coefficients of a dual algebra ${\mathfrak{\tilde
g}}$ and $v_{i}$ are left-invariant fields on the Lie group $G$. The Lie algebras
${\mathfrak{g}}$ and ${\mathfrak{\tilde g}}$ then define the \dd{} as a connected Lie group
whose Lie algebra ${\mathfrak{d}}$ can be decomposed into the pair of subalgebras
${\mathfrak{g}},\tilde{\mathfrak g}$ {that are} maximally isotropic with respect to a symmetric
ad-invariant nondegenerate bilinear form $<.,.>$ on ${\mathfrak{d}}$.

\dd s enable {us} to construct the tensor $F$ satisfying (\ref{dc}). The general solution
of this equation is
    \be\label{met}
        F_{\mu\nu}(x)=e_{\mu}^{a}(g(x))E_{ab}(g(x))e_{\nu}^{b}(g(x)),
    \ee
where $e_{\mu}^{a}(g(x))$ are components of right-invariant forms
$e_{\mu}^{a}=((dg)g^{-1})_{\mu}^{a}$ expressed in coordinates $x^\mu$ on the group $G$,
\be\label{metr}
        E(g)=[E_{0}^{-1}+\Pi(g)]^{-1},
    \ee
$E_{0}$ is a constant {non-singular} matrix, $\Pi(g)$ is given by the formula
    \be
        \Pi(g)=b(g).a(g)^{-1}=-\Pi^{t}(g),
    \ee
and matrices $a(g),b(g),d(g)$ are given by the adjoint representation of the Lie subgroup
$G$ on the Lie algebra of the Drinfeld double in the mutually dual bases 
    \be
        Ad(g)^{T}=\left(
                \begin{array}{cc}
                    a(g) & 0 \\
                    b(g) & d(g) \\
                \end{array}
      \right).
    \ee
If $\wt G$ is Abelian, then $\Pi(g)=0$.

Under the condition (\ref{dc}), the field equations for the \sm{} can be rewritten as the
\eqn
    \be \label{deqn}
        <(\partial_{\pm}l)l^{-1},\varepsilon^{\pm}>=0
    \ee
for the mapping $l$ from the worldsheet in $\mathbb{R}^2$ into the Drinfeld double $D$,
where the subspaces $\varepsilon^{+}=Span(T_{i}+E_{0,ij}{\tilde T}^{j})$,
$\varepsilon^{-}=Span(T_{i}-E_{0,ji}{\tilde T}^{j})$ are orthogonal w.r.t. $<,>$ and span
the whole Lie algebra $\mathfrak d$, and $\{T_{i}\}$, $\{{\tilde T}^{j}\}$ are the the mutually dual
bases of ${\mathfrak{g}}$ and $\tilde{\mathfrak g}$.

Due to Drinfeld, there exists a unique decomposition (at least in the vicinity of the unit element of $D$) of an
arbitrary element $l$ of $D$ as a product of elements
from $G$ and $\tilde G$. Solutions of \eqn{} (\ref{deqn}) and solution of the \eqn s of
motion for the \sm {} $\phi^\mu(\tau,\sigma) = x^\mu( g(\tau,\sigma))$ are related by
    \be
        l(\tau,\sigma)=g(\tau,\sigma){\tilde{h}}(\tau,\sigma)\in D,
    \ee
where ${\tilde {h}}\in {\tilde{G}}$ fulfills the equations \be ((\partial_\tau \tilde
h).\tilde h\-1)_j = 
-v^\lambda_j\left[G_{\lambda\nu}\partial_\sigma\phi^\nu+B_{\lambda\nu}\partial_\tau\phi^\nu\right]
\label{btp},\ee\be ((\partial_\sigma \tilde h).\tilde h\-1)_j =
-v^\lambda_j\left[G_{\lambda\nu}\partial_\tau\phi^\nu+B_{\lambda\nu}\partial_\sigma\phi^\nu\right]
 \label{btm}.\ee
{and $v_j^\lambda$ are components of the left-invariant fields on $G$}.

The dual model can be obtained by the exchange \be \label{duality}
G\leftrightarrow{\tilde G},\ \ \ {\cg}\leftrightarrow
{\mathfrak{\tilde g}},\ \ \ \Pi(g)\leftrightarrow
{\tilde{\Pi}(\tilde g)},\ \ \ E_{0}\leftrightarrow
E_{0}^{-1}.\ee

The relation between the solution $\phi^\mu(\tau,\sigma) $ of the \eqn s of motion of the
sigma model given by $F$ {and the solution} $\wt \phi^\mu(\tau,\sigma) $ of the model given
by $\wt F$ follows from two possible decompositions of elements $l$ of the Drinfeld double
\be \label{lghtgth}
         g(\tau,\sigma)\tilde h(\tau,\sigma)=\tilde g(\tau,\sigma)h(\tau,\sigma)
   , \ee where $g, h \in G,\ \wt g, \wt h\in \wt G$. The map $ \wt h:\real^2\rightarrow \tilde G$ that we need
for
this transformation {is the solution to} the equations  
(\ref{btp},\ref{btm}).

The equation (\ref{lghtgth}) {then} defines the \pl {} transformation between the solution
of the \eqn s of motion {of the original \sm {} and its dual. Its} application may be very
complicated. To use it for solving the dual model {the following} three steps must be done:
\begin{itemize}
\item One has to know the solution $\phi^\mu(\tau,\sigma)$ of the \sm{} given by $F$.
\item {Given} $\phi^\mu(\tau,\sigma)$, one has to find $\wt h(\tau,\sigma)$
i.e. solve the system of PDEs (\ref{btp},\ref{btm}).
\item {Given $l(\tau,\sigma)= g(\tau,\sigma)\tilde h(\tau,\sigma)\in D$, one has to find the dual decomposition
$l(\tau,\sigma)=\tilde g(\tau,\sigma)h(\tau,\sigma)$, where $\wt g(\tau,\sigma)\in \wt G$,
$h(\tau,\sigma)\in G$}.
\end{itemize}
In the {following sections we shall apply these three steps of} the \pl {} transformation
to solve the equations of motion for strings in backgrounds dual to the homogeneous
plane--parallel wave metric. {We will start with the description of a particular plane wave
background and recall the solution of its classical equations of motion.}

\section{Homogeneous plane--parallel wave metrics}
Homogeneous plane--parallel wave in $d+2$ dimensions is generally defined by the metric of
the following form \cite{papa},\cite{blaulough},\cite{blauffpapa}
    \be \label{homogeneousmetric}
        ds^{2}=2dudv-K_{ij}(u)x^{i}x^{j}du^{2}+d\vec{x}^{2},
    \ee
where $d\vec{x}^{2}$ is the standard metric on Euclidean space ${\bf{E}}^{d}$ and
$\vec{x}\in{\bf{E}}^{d}.$ The form of this metric seems to be simple, but explicit solution
of sigma models can be complicated. {Therefore the authors of \cite{papa} restricted
themselves to the special case of isotropic homogeneous plane wave metric}
\begin{equation}\label{Aij}
K_{ij}(u)=\lambda(u)\delta_{ij},\ \
    \lambda(u)=\frac{k}{u^{2}},\quad k=\nu(1-\nu)=const.> 0.
\end{equation}
In the following we shall investigate the case $d=2$, i.e. the dimension of the spacetime
is four. This seems to be the simplest physically interesting background\footnote{The study
of higher-dimensional cases would mean that we would have to look for $(d+2)$-dimensional
subalgebras of the algebra of Killing vectors of the dimension $2+\half d (d+3)$.}. The
metric tensor in the (Brinkman) coordinates $(u,v,x,y)$ then has components
    \be\label{FF}
        G_{\mu\nu}(u,v,x,y)=\left(
          \begin{array}{cccc}
            \frac{-k(x^{2}+y^{2})}{u^{2}} & 1 & 0 & 0 \\
            1 & 0 & 0 & 0 \\
            0 & 0 & 1 & 0 \\
            0 & 0 & 0 & 1 \\
          \end{array}
        \right).
    \ee
This metric is not flat, but its Gaussian curvature vanishes. It has a
singularity in $ u=0$, and does not satisfy the Einstein \eqn s but
the conformal invariance condition equations for vanishing of the
$\beta$ function
\begin{eqnarray}
\label{bt1} 0 & = & R_{\mu\nu}-\bigtriangledown_\mu\bigtriangledown_\nu\Phi-
\frac{1}{4}H_{\mu\kappa\lambda}{H_\nu}^{\kappa\lambda},
\\
 \label{bt2} 0 & = & \bigtriangledown^\mu\Phi H_{\mu\kappa\lambda}+\bigtriangledown^\mu H_{\mu\kappa\lambda}\,,
\\
\label{bt3} 0 & = & R-2\bigtriangledown_\mu\bigtriangledown^\mu\Phi-
\bigtriangledown_\mu\Phi\bigtriangledown^\mu\Phi-
\frac{1}{12}H_{\mu\kappa\lambda}H^{\mu\kappa\lambda},
\end{eqnarray}
where the covariant derivatives $\bigtriangledown_k$, the Ricci tensor $R_{\mu\nu}$ and the scalar
curvature $R$ are calculated from the metric $G_{\mu\nu}$ that is also used for lowering
and raising indices. The torsion $H$ as well as its potential $B$  in this case vanishes, and
the dilaton field is
\begin{equation}\label{dilatonhomogeneous} \Phi= \Phi_0 - c\,u+
2\nu(\nu-1)\ln\, u.
\end{equation}
The metric admits {symmetries generated by} the following Killing vectors
   \begin{align}
\label{killingshomogeneous}
\nonumber        K_{1}&=\partial_{v},\\
\nonumber        K_{2}&=u^{\nu}\partial_{x}-\nu
        u^{\nu-1}x\partial_{v},\\
\nonumber        K_{3}&=u^{\nu}\partial_{y}-\nu
        u^{\nu-1}y\partial_{v},\\
        K_{4}&=u^{1-\nu}\partial_{x}-(1-\nu)u^{-\nu}x\partial_{v},\\
\nonumber       K_{5}&=u^{1-\nu}\partial_{y}-(1-\nu)u^{-\nu}y\partial_{v},\\
\nonumber        K_{6}&=u\partial_{u}-v\partial_{v},\\\nonumber
        K_{7}&=x\partial_{y}-y\partial_{x}.
        \end{align}
One can easily check that the Lie algebra spanned by these vectors is the semidirect sum $
{\mathcal S}\ltimes {\mathcal N}$ {of} ${\mathcal S}=Span[K_6,K_7]$ and an ideal ${\mathcal
N}=Span[K_1, K_2, K_3, K_4, K_5]$. The algebra ${\mathcal S}$ is Abelian and its generators can be
interpreted as dilation in $u,v$ and rotation in $x,y$. Generators of the algebra ${\mathcal
N}$ commute as the two-dimensional Heisenberg algebra with the center $K_1$.

The equations of motion for
$\phi^\mu(\tau,\sigma)=(U(\tau,\sigma),V(\tau,\sigma),X(\tau,\sigma),Y(\tau,\sigma))$ are
given by the sigma model  action \eqref{sma}. They read
\begin{eqnarray*}
  (\partial_\sigma^2-\partial_\tau^2)U &=& 0 ,\\
  (\partial_\sigma^2-\partial_\tau^2)X +\frac{k}{U^2}\partial_aU\partial^aU X&=& 0,\\
  (\partial_\sigma^2-\partial_\tau^2)Y +\frac{k}{U^2}\partial_aU\partial^aU Y&=&
  0,
\end{eqnarray*}
\[(\partial_\sigma^2-\partial_\tau^2)V+\frac{k}{U^3}\,\partial_aU\partial^aU(X^2+Y^2)-
  \frac{2k}{U^2}\partial_aU(\partial^aX X+\partial^aY Y)= 0. \]

{The plane wave metric \eqref{FF} allows to adopt the light-cone
gauge}\begin{equation}\label{lightcg}
    U(\tau,\sigma)=\kappa \tau,\ \ \kappa:=2\alpha'p^u,
\end{equation}
{in which the equations of motion simplify and acquire the form}
\begin{eqnarray}
  (\partial_\sigma^2-\partial_\tau^2)X -\frac{k}{\tau^2}X&=& \label{eqnforX} 0,\\
  (\partial_\sigma^2-\partial_\tau^2)Y -\frac{k}{\tau^2}Y&=& 0 \label{eqnforY},\\
 (\partial_\sigma^2-\partial_\tau^2)V-\frac{k}{\kappa\tau^3}(X^2+Y^2)+
  \frac{2k}{\kappa\tau^2}(\partial_\tau X X+\partial_\tau Y Y)&=& 0\label{eqnforV} .
 \end{eqnarray}
Solution of the \eqn s (\ref{eqnforX}), (\ref{eqnforY}) that for $\tau\rightarrow \infty$
tends to the free string {solution} was given in \cite{papa} as
\begin{eqnarray}
 X^i(\sigma,\tau ) &=& x_0^i(\tau ) + {i\over 2} \sqrt{2
\alpha '} \sum_{n=1}^\infty {1\over n}
\bigg[X^i_n(\tau,\sigma)-X^{i*}_n(\tau,\sigma)\bigg]
,\end{eqnarray} {where $i=2,3,\ X=X^2,Y=X^3, \nu =  {1\over 2} (1 + \sqrt{1-4 k}),$ the
zero modes are}
\begin{equation}
x_0^i(\tau )=  {1\over \sqrt{2\nu-1} }\big( \tilde x^i \,
\tau^{1-\nu }
 + 2\alpha' \tilde p^i \, \tau^{\nu } \big),\ \text{for } k\neq {1\over 4},
\end{equation}
\begin{equation} \label{k14}x_0^i(\tau )= \sqrt{\tau }( \tilde x^i  + 2\alpha' \tilde p^i \ \log \tau ),\ \text{for } k={1\over 4}. \end{equation}
{The higher modes are expanded as}
\begin{equation}X^i_n(\tau,\sigma)=Z(2n\tau ) \big( \alpha_n^i e^{2 in \sigma} +
\tilde \alpha_n^i e^{-2 in \sigma} \big),
\end{equation}\begin{equation}
Z(2n\tau ):= e^{-i {\pi\over 2} \nu } \sqrt{\pi n \tau }\ H^{(2)}_{\nu -{1\over 2}}(2n\tau
),
\end{equation}and  $H^{(2)}_{\nu -{1\over 2}}$ is the Hankel function of the second kind
\begin{equation} H^{(2)}_{\nu -{1\over 2}}(t)=\big[ J_{\nu -{1\over 2} }(t) - i\, Y_{\nu -{1\over 2} }(t ) \big]\ . \ \ \
\end{equation}

{Being} interested in the string solutions of the sigma model, we have to add supplementary
string conditions
\begin{equation}\label{stringcond}
    \partial_a\Phi^\mu G_{\mu\nu}(\Phi)\partial_b\Phi^\nu=e^\omega\eta_{ab}
\end{equation} that for $\eta=diag (-1,1)$ and in the light-cone gauge read
\begin{equation}\label{stringcond1}
    \kappa\partial_\sigma V
 +
  \partial_\tau X\partial_\sigma X +\partial_\tau Y\partial_\sigma Y
  =0,
\end{equation}
\begin{equation}\label{stringcond2}
    2\kappa\partial_\tau V-
  \frac{k}{\tau^2}(X^2+Y^2)+
  \partial_\tau X\partial_\tau X+\partial_\sigma X\partial_\sigma X
   +\partial_\tau Y\partial_\tau Y+\partial_\sigma Y\partial_\sigma Y  =
   0.
\end{equation}
Compatibility of these two equations is guaranteed by the equations of motion for $X$ and
$Y$ (\ref{eqnforX}) and (\ref{eqnforY}), so that
\begin{equation}\label{solnV}
    V(\tau,\sigma)=v(\tau)-\frac{1}{\kappa}\int d\sigma (\partial_\tau X\partial_\sigma X +\partial_\tau Y\partial_\sigma
    Y),
\end{equation}
{where $v(\tau)$ is an arbitrary function. The} \eqn {} (\ref{eqnforV}) is solved by
(\ref{solnV}) provided that the functions $X,Y$ satisfy  (\ref{eqnforX}) and
(\ref{eqnforY}).

\section{Data for construction of dual backgrounds }\label{homogeneous} As explained in section \ref{secPLT}, dualizable metrics
are constructed by virtue of the \dd. {To get the metric (\ref{FF}), the Lie algebra of the
Drinfeld double can be composed from the four-dimensional Lie subalgebra of the isometry
algebra of Killing vectors \eqref{killingshomogeneous} and four-dimensional Abelian
algebra\footnote{The equation $(\ref{dc})$ is then trivially fulfilled.}}. Moreover, the
four-dimensional subgroup of isometries must act freely and transitively \cite{klise} on
the Riemannian manifold $M$ where the metric is defined, so that $M\approx G$. These
four-dimensional subgroups were found in \cite{hlatur3} as
$\mathfrak{g}_1=Span\{K_1,K_2,K_5,K_6\}$  or $\mathfrak{g}_2=Span\{K_1,K_2,K_3,K_6+\rho\,
K_7\}$ with non-vanishing commutation relations
\begin{align}
\nonumber        [K_{6},K_{1}]&=K_{1},\\
\label{cr2}        [K_{6},K_{2}]&=\nu\, K_{2},\\
\nonumber        [K_{6},K_{5}]&=(1-\nu)\, K_{5},
\end{align}
and
        \begin{align}
    \nonumber    [K_{6}+\rho K_7,K_{1}]&=K_{1},\\
  \label{cr1}      [K_{6}+\rho K_7,K_{2}]&=\nu\, K_{2}-\rho\ K_3,\\
  \nonumber        [K_{6}+\rho K_7,K_{3}]&=\nu\, K_{3}+\rho\ K_2
    \end{align}
respectively, where {the parameter $\nu$ was given in \eqref{Aij}} and $\rho$ is an
arbitrary real.

In the following we shall find metrics dual to (\ref{FF}) constructed from the \dd s
$\mathfrak d=\mathfrak g\oplus\mathfrak a$ where $\mathfrak g$ is  $\mathfrak{g}_1$ or
$\mathfrak{g}_2$ and $\mathfrak a$ is four-dimensional Abelian algebra.

To get the matrix $E_0$, the metric (\ref{FF}) must be transformed to coordinates
$x^{1},x^{2},x^{3},x^{4}$ used for parametrization of the group elements identified with
points of the manifold. {Choosing the parametrization of the elements of the group $G$ as
    \be\label{gel}
        g(x)=e^{x^{1}T_{1}}e^{x^{2}T_{2}}e^{x^{3}T_{3}}e^{x^{4}T_{4}},
    \ee
where $T_{1},T_{2},T_{3},T_{4}$ are generators of the group $G$, the matrix $E_0$ was found
in \cite{hlatur3} to be}
\begin{equation}\label{E0cite}
    E_0=\left(
\begin{array}{cccc}
 0 & 0 & 0 & 1 \\
 0 & 1 & 0 & 0 \\
 0 & 0 & 1 & 0 \\
 1 & 0&0 &0
\end{array}\right).
\end{equation}
The dual metrics on the group $\tilde G$ with elements \be
        \wt g(\wt x)=e^{\wt x_{1}\wt T^{1}}e^{\wt x_{2}\wt T^{2}}e^{\wt x_{3}\wt T^{3}}e^{\wt x_{4}\wt T^{4}},
    \ee
as well as classical solutions of their dual sigma model
\begin{equation}\label{solutionofdual1256}
    \wt g(\tau,\sigma)=e^{\wt x_{1}(\tau,\sigma)\wt T^{1}}e^{\wt x_{2}(\tau,\sigma)\wt T^{2}}
    e^{\wt x_{3}(\tau,\sigma)\wt T^{3}}e^{\wt x_{4}(\tau,\sigma)\wt T^{4}},
\end{equation} can be then obtained by the method described
in the Sec. \ref{secPLT}. {We will use it in the following two sections.}

\section{Classical strings in the dual background obtained from
$\mathfrak{g}_1$ } Let us first consider the group generated by the Lie algebra
$\mathfrak{g}_1=Span[T_1, T_2, T_3, T_4]$ with commutation relations (cf.
(\ref{cr2}))\begin{eqnarray}
   \nonumber
        [T_{4},T_{1}]&=&T_{1},\\ \label{1256} [T_{4},T_{2}]&=&\nu\, T_{2},\\ \nonumber
        [T_{4},T_{3}]&=&(1-\nu)T_{3}.
\end{eqnarray}
The transformation between group coordinates $x^{1},x^{2},x^{3},x^{4}$ defined by
(\ref{gel}) and geometrical coordinates $u,v,x,y$ on $M$ can be obtained by comparing the
left-invariant vector fields on the group $G$ and Killing vectors of the metric
(\ref{FF}). One gets\ba \nonumber
        u&=&e^{x^{4}},\\ \label{tfncoorshomogeneous2}
        v&=&\frac{1}{2}\,[2x^{1}-\nu(x^{2})^{2}-(1-\nu)(x^{3})^{2}]e^{-x^{4}},\\\nonumber
        x&=&x^{2},\\\nonumber
        y&=&x^{3}.
    \ea
The metric (\ref{FF}) expressed in the group coordinates then acquires the form \be
G_{\mu\nu}(x^\lambda)=\left(
\begin{array}{cccc}
 0 & 0 & 0 & 1 \\
 0 & 1 & 0 & -\nu  x^2 \\
 0 & 0 & 1 & (\nu -1) x^3 \\
 1 & -\nu  x^2 & (\nu -1) x^3 & \nu ^2 (x^2)^2+(\nu -1)^2 (x^3)^2-2 x^1
\end{array}
\right).\ee

The group $\wt G$ is Abelian and the right-hand sides of the \eqn s (\ref{btp},\ref{btm})
are invariant w.r.t coordinate transformations. {That is why in the Brinkman coordinates
$u,v,x,y$ we can use just $K_1, K_2, K_5, K_6$ as the left-invariant fields on $G$. }The
equations (\ref{btp},\ref{btm}) for $\tilde h$ then read
\begin{equation}\label{atau1}
    \partial_\tau\tilde h=-\left(
\begin{array}{c}
 0 \\
 (\kappa  \tau )^{\nu }\partial_{\sigma} X \\
 (\kappa  \tau )^{1-\nu } \partial_{\sigma} Y \\
 \kappa  \tau\,  \partial_{\sigma} V
\end{array}
\right),
\end{equation}
\begin{equation}\label{asigma1}
   \partial_\sigma\tilde h= -\left(
\begin{array}{c}
 \kappa  \\
 (\kappa  \tau )^{\nu } \left(\partial_{\tau} X-\frac{\nu}{\tau}  X\right)\\
 \kappa  (\kappa  \tau )^{-\nu } \left(\tau  \partial_{\tau} Y+(\nu -1) Y\right) \\
 \frac{\nu(\nu-1)}{\tau }( X^2+ Y^2)+\kappa  \tau \partial_{\tau} V-\kappa   V\end{array}
\right),
\end{equation}
and are solved by
\begin{eqnarray}
\label{lcsoln}  \tilde h_1 &=&c_1-\kappa\sigma  ,\\
  \tilde h_2 &=& c_2 -(\kappa\tau)^\nu\int d\sigma(\partial_\tau X-\frac{\nu}{\tau}X)\label{h2},\\
  \tilde h_3 &=& c_3 -(\kappa\tau)^{(1-\nu)}\int d\sigma(\partial_\tau Y-\frac{1-\nu}{\tau}Y)\label{h3},\\
  \tilde h_4 &=& c_4 +\int d\sigma[\frac{\nu(1-\nu)}{\tau}(X^2+Y^2)+\kappa(V-\tau\partial_\tau
  V)]\\ \nonumber
  &=&\int d\sigma\left[\frac{\nu(1-\nu)}{2\tau}(X^2+Y^2)+\frac{\tau}{2}\left[(\partial_\tau X)^2+(\partial_\sigma X)^2 +
  (\partial_\tau Y)^2+(\partial_\sigma Y)^2\right]\right]\\
  &&
  -\int d\sigma\int d\sigma'(\partial_\tau X\partial_\sigma X +\partial_\tau Y\partial_\sigma
  Y)+c_4+\kappa\,\sigma\, v(t),
\end{eqnarray}
{where $c_i, i=1,\ldots,4$ are arbitrary constants.}

The dual tensor on the group $\wt G$ is constructed by the procedure explained in the
section \ref{secPLT}, namely by using (\ref{met}),(\ref{metr}), (\ref{duality})
\begin{equation}\label{dualhomogenoeous2}
\tilde  F_{\mu\nu}(\tilde{x})=   \left(
\begin{array}{cccc}
 \frac{\nu ^2 \tilde{x}_2^2+\left(1-\nu\right)^2\tilde{x}_3^2}{\tilde{x}_1^2-1} & \frac{\nu\,  \tilde{x}_2}{1-\tilde{x}_1} &
 \frac{\left(1-\nu\right)\,
   \tilde{x}_3}{1-\tilde{x}_1} & \frac{1}{1-\tilde{x}_1} \\
 -\frac{\nu\,  \tilde{x}_2}{\tilde{x}_1+1} & 1 & 0 & 0 \\
 \frac{\left(\nu-1\right)\,  \tilde{x}_3}{\tilde{x}_1+1} & 0 & 1 & 0 \\
 \frac{1}{\tilde{x}_1+1} & 0 & 0 & 0
\end{array}
\right).
\end{equation}
Even though it is not symmetric, its torsion is zero. It satisfies the conformal invariance
conditions (\ref{bt1})--(\ref{bt3}) with the dilaton field \cite{hlatur3}
\begin{equation}\label{dualdilatonhomogeneous3} \tilde \Phi= \tilde \Phi_0 +
C \ln\left(\frac{\tilde{x}_1-1}{\tilde{x}_1+1}\right) +(\nu-1-\nu^2)\ln(1-\tilde{x}_1^2)
\end{equation}
that we rederive later.

To obtain the solution of the \sm {} on $\wt G$ given by \be\label{sma2}
        S_{\tilde  F}[\wt x]=-\int d\sigma_+d\sigma_-\,(\partial_{-}\wt x^{\mu}\tilde  F_{\mu\nu}(\wt x)\partial_{+}\wt x^{\nu}),
    \ee
we {have to} solve the equation (\ref{lghtgth}) for $\wt x_{j}$, where
\begin{eqnarray}
\label{ght}
  g=e^{x^{1}T_{1}}e^{x^{2}T_{2}}e^{x^{3}T_{3}}e^{x^{4}T_{4}},&&\  \wt h=e^{\wt h_{1}\wt T^{1}}e^{\wt h_{2}\wt T^{2}}e^{\wt h_{3}\wt T^{3}}e^{\wt h_{4}\wt T^{4}}
          ,\\ \label{gth}
 \wt g=e^{\wt x_{1}\wt T
 ^{1}}e^{\wt x_{2}\wt T^{2}}e^{\wt x_{3}\wt T^{3}}e^{\wt x_{4}\wt T^{4}}, && h=e^{h^{1} T
 _{1}}e^{h^{2}T_{2}}e^{h^{3} T_{3}}e^{h^{4} T_{4}}.
\end{eqnarray}
To accomplish this, we can use a representation $r$  of the semi-abelian \dd{}  in the form of
block matrices $(dim {\mathfrak g}+1)\times (dim {\mathfrak g}+1)$, such that
\begin{equation}\label{Drep}
   r(g)= \left(
\begin{array}{cc}
Ad\, g & 0 \\
 0& 1
\end{array}
\right),\ \ \   r(\wt h)= \left(
\begin{array}{cc}
{\bf 1}& 0 \\
 v(\wt h)& 1
\end{array}
\right),
\end{equation}
where $v(\wt h)=(\wt h_1,\ldots,\wt h_{dim\mathfrak g})$. From the equation (\ref{lghtgth}) we then get
\begin{equation}\label{replgh}
   r(l)=r(g\wt h)= \left(
\begin{array}{cc}
Ad\, g & 0 \\
 v(\wt h)& 1
\end{array}
\right) =r(\wt g  h)= \left(
\begin{array}{cc}
Ad\, h & 0 \\
 v(\wt g)\cdot(Ad\, h )& 1
\end{array}
\right). \end{equation} If the adjoint representation of the Lie algebra $\mathfrak g$ is
faithful, which is the case of (\ref{1256}), then the representation $r$ of the \dd {} is
faithful as well and the relation (\ref{replgh}) gives a system of \eqn s for $\tilde x_j$
and $h^j$.

Plugging (\ref{ght}) and (\ref{gth}) into (\ref{replgh}) together with the the adjoint
representation of the Lie algebra (\ref{1256}) we get the solution of (\ref{lghtgth}) {in
the form}
\begin{eqnarray}
  h^j&=& x^j \label{tx0} ,\\
  \tilde x_1 &=& \e^{-x^4}\tilde h_1 \label{tx1},\\ \label{tx2}  \tilde x_2 &=& \e^{-\nu x^4}\tilde h_2,\\
  \tilde x_3 &=& \e^{(\nu-1)x^4}\tilde h_3 \label{tx3} ,\\
  \tilde x_4 &=& \e^{-x^4}x^1\tilde h_1+\nu\e^{-\nu x^4}x^2\tilde h_2
  + (1-\nu)\e^{(\nu-1)x^4}x^3\tilde h_3 +\tilde h_4. \label{tx4}
\end{eqnarray}
{The expressions for $ \tilde x_1,\tilde x_2,\tilde x_3$ are very simple and it can be
checked that combining them with (\ref{tfncoorshomogeneous2}) they} {\em give explicit
solution of the equations of motion of the \sm {} on $\wt G$ in the background}
(\ref{dualhomogenoeous2}). Namely, inserting the light-cone gauge solution (\ref{lightcg})
into (\ref{tfncoorshomogeneous2}) and (\ref{tx1}) we get from (\ref{lcsoln})
\begin{equation}\label{tx1u}
    \wt X_1(\tau,\sigma)= \frac{c_1-\kappa\sigma }{\kappa \tau}
\end{equation} that solves the dual \eqn {} of motion
\begin{equation}\label{eqnfortx1}
\frac{\delta S_{\wt F}}{\delta \wt X_4} =\left(\partial_{\sigma}^2\wt
X_1-\partial_{\tau}^2\wt X_1\right)\left( 1-\wt
   X_1^2\right)+
   2\, \wt
   X_1\left((\partial_{\sigma}\wt X_1)^2-(\partial_{\tau}\wt X_1)^2\right) =0.
   \end{equation}
Other two \eqn s then reduce to
\begin{eqnarray}\label{eqnfortx2}
 \frac{\delta S_{\wt F}}{\delta \wt X_2} = (\partial_\sigma^2-\partial_\tau^2)\wt X_2 +\frac{\nu(1+\nu)}{\tau^2}\wt X_2&=&0,\\ \label{eqnfortx3}
\frac{\delta S_{\wt F}}{\delta \wt X_3} =  (\partial_\sigma^2-\partial_\tau^2)\wt X_3
+\frac{(\nu-2)(\nu-1)}{\tau^2}\wt X_3&=&0,
\end{eqnarray} solved in agreement with (\ref{tx2},\ref{tx3}) and (\ref{tfncoorshomogeneous2},\ref{h2},\ref{h3}) as
$$
\nonumber \wt X_2(\tau,\sigma) =c_2(\kappa\tau)^{-\nu}+ \int d\sigma(\frac{\nu}{\tau}X-\partial_\tau X)\\
,$$
$$\wt X_3(\tau,\sigma)=c_3(\kappa\tau)^{\nu-1}+ \int d\sigma(\frac{1-\nu}{\tau}Y-\partial_\tau
Y).$$ For $\nu\neq\frac{1}{2}$ we have
\begin{equation}
 \label{X2} \wt X_2(\tau,\sigma)=( {c_2}\,
\kappa^{-\nu }+\sqrt{2 \nu -1}\, \tilde x\,\sigma)\tau^{-\nu}+\Sigma_2^x(\tau,\sigma)
,\end{equation}
\begin{equation} \label{X3}\wt X_3(\tau,\sigma)=({c_3}\, \kappa^{\nu-1}-\sqrt{2 \nu -1}\
2\,\tilde p^y\alpha'\,\sigma)\tau^{\nu-1}+\Sigma_3^y(\tau,\sigma),
\end{equation}
where
\begin{eqnarray}
\Sigma_2^j(\tau,\sigma)&=& \half\sqrt{2\alpha'}\,\sum_{n=1}^\infty\frac{1}{n}\,\left[W(2 n
\tau)
  (\alpha_{n}^j e^{2 i n \sigma}-\tilde\alpha_{n}^j e^{-2 i n \sigma})\right.\nonumber\\&&\label{Sigma2}\left.
  +\ W^*(2 n \tau)
  (\alpha_{-n}^j e^{-2 i n \sigma}-\tilde\alpha_{-n}^j e^{2 i n \sigma})\right],\\
\Sigma_3^j(\tau,\sigma)&=&\half\sqrt{2\alpha'}\,\sum_{n=1}^\infty\frac{1}{n}\,\left[\wt W(2
n \tau)
  (\alpha_{n}^j e^{2 i n \sigma}-\tilde\alpha_{n}^j e^{-2 i n \sigma})\ \right.\nonumber\\&&\label{Sigma3}\left.
  +\ \wt W^*(2 n \tau)
  (\alpha_{-n}^j e^{-2 i n \sigma}-\tilde\alpha_{-n}^j e^{2 i n \sigma})\right],
\end{eqnarray}
and\begin{eqnarray*} W(2 n
\tau)&=&\frac{1}{2n}\left(\frac{\nu}{\tau}Z(2n\tau)-\partial_\tau Z(2n\tau)\right)=e^{
-\frac{i \pi \nu }{2}}\sqrt{\pi n \tau}\,\
  H_{\nu +\frac{1}{2}}^{(2)}(2 n \tau),\\
\wt W(2 n \tau)&=&\frac{1}{2n}\left(\frac{1-\nu}{\tau}Z(2n\tau)-\partial_\tau
Z(2n\tau)\right)=-\,e^{ -\frac{i \pi  \nu }{2}}\sqrt{\pi n \tau}\,\
  H_{\nu -\frac{3}{2}}^{(2)}(2 n \tau).\end{eqnarray*}
For $\nu=\frac{1}{2}$ the expressions are a bit different and can be derived from
\eqref{k14}.

Finally, from the expression (\ref{tx4}) we get
$$\wt X_4(\tau,\sigma) =c_4-\half\frac{\sigma + c_1}{ \tau}\Big(2 \kappa\tau\, V
 +\nu X^2+(1-\nu) Y^2\Big)+\nu X\tilde X_2+
  (1-\nu)Y\tilde X_3$$
$$+\int d\sigma[\frac{\nu(1-\nu)}{\tau}(X^2+Y^2)+\kappa(V-\tau\partial_\tau
  V)]
$$ that solves the last equation of the dual sigma model.

\subsection{Killing vectors for the dual metric  and its
plane--parallel wave form}\label{42} The purpose of this section is to show that the metric
corresponding to the dual tensor (\ref{dualhomogenoeous2}) is again a plane--parallel wave.

In this case the relevant part of the dual tensor is only its symmetric part because the
torsion $H=d B$ vanishes and does not influence neither the \eqn s of motion nor the
$\beta$ \eqn s (\ref{bt1})--(\ref{bt3}). The dual metric calculated from
(\ref{dualhomogenoeous2}) is
\begin{equation}\label{dualmetric1}
\tilde  G_{\mu\nu}(\tilde{x})=   \left(
\begin{array}{cccc}
 \frac{\nu ^2 {\tilde  x_2}{}^2+(1-\nu )^2{\tilde  x_3}{}^2}{{\tilde  x_1}{}^2-1}
 & \frac{\nu\,  {\tilde  x_1}{} {\tilde  x_2}{}}{1-{\tilde  x_1}{}^2} &
   \frac{(1-\nu) {\tilde  x_1}{} {\tilde  x_3}{}}{{1-\tilde  x_1}{}^2} & \frac{1}{1-{\tilde  x_1}{}^2} \\
 \frac{\nu\,  {\tilde  x_1}{} {\tilde  x_2}{}}{1-{\tilde  x_1}{}^2} & 1 & 0 & 0 \\
 \frac{(1-\nu) {\tilde  x_1}{} {\tilde  x_3}{}}{{1-\tilde  x_1}{}^2} & 0 & 1 & 0 \\
 \frac{1}{1-{\tilde  x_1}{}^2} & 0 & 0 & 0
\end{array}
\right).
\end{equation}
This metric for $\nu\neq 1,0$,  i.e. $k\neq 0$ has just five-dimensional algebra
{generated by} Killing vectors
\begin{eqnarray}
\label{dualkil1}
  \tilde K_1&=& P_\nu(\tilde  x_1)\frac{\partial}{\partial \tilde x_2}
  +\tilde x_2[(\nu+1)P_{\nu+1}(\tilde  x_1)-\tilde  x_1(1+2\nu)P_\nu(\tilde  x_1)]\frac{\partial}{\partial \tilde x_4} ,\\
  \tilde K_2&=&  Q_\nu(\tilde  x_1)\frac{\partial}{\partial \tilde x_2}
  +\tilde  x_2[(\nu+1)Q_{\nu+1}(\tilde  x_1)-\tilde  x_1(1+2\nu)Q_\nu(\tilde  x_1)]\frac{\partial}{\partial \tilde x_4} ,\\
 \tilde K_3&=& P_{\nu-2}(\tilde  x_1)\frac{\partial}{\partial \tilde x_3}
  +\tilde  x_3(\nu-1)P_{\nu-1}(\tilde  x_1)\frac{\partial}{\partial \tilde x_4} ,\\
   \tilde K_4&=& Q_{\nu-2}(\tilde  x_1)\frac{\partial}{\partial \tilde x_3}
  +\tilde  x_3(\nu-1)Q_{\nu-1}(\tilde  x_1)\frac{\partial}{\partial \tilde x_4} ,\\
  \tilde K_5&=&-\frac{\partial}{\partial \tilde x_4},\label{dualkil5}\end{eqnarray}
  where $P_\nu$ and $Q_\nu$ are Legendre functions of the first and second kind.
The commutators close to the Heisenberg algebra with the central element $\tilde K_5$
\begin{equation}\label{tilkil1}
    [\tilde K_1,\tilde K_2]=\tilde K_5,\ \ [\tilde K_3,\tilde K_4]=\tilde K_5
\end{equation}  due to the identity
\begin{equation}\label{identity}
   P_{1+\nu}(z) Q_\nu(z) - Q_{1+\nu}(z) P_\nu(z)  = 1/(1+\nu).
\end{equation}

{The number of the Killing vectors as well as the fact that they close} to the Heisenberg
algebra suggests (cf.  \cite{blaulough}) that this metric might be brought to the form of
plane--parallel wave. Trying to rewrite the Killing vectors
(\ref{dualkil1})--(\ref{dualkil5}) to new (Rosen) coordinates $z,w,y^1,y^2$ {such that the
Killing vectors acquire the form} \cite{blauffpapa}
\begin{equation}\label{kilbffpp}
    e_+=\frac{\partial}{\partial w},\   e_i=\frac{\partial}{\partial y^i},
    \    e_i^*=y^i\frac{\partial}{\partial w}-\Gamma^{ij}(z)\frac{\partial}{\partial y^i},\
    i,j\in\{1,2\},
\end{equation}
we can get the transformation
\begin{eqnarray}
 \nonumber  \tilde x_1 &=& z ,\\
 \nonumber  \tilde x_2  &=& y^1 P_\nu(z),\\
 \label{XiInRosen}  \tilde x_3  &=& y^2 P_{\nu-2}(z),\\
 \nonumber \tilde x_4 &= & -w + \half\Big[(y^1)^2 P_{\nu }(z) \Big((\nu +1) P_{\nu +1}(z)-z(1+2  \nu) P_{\nu
 }(z)\Big)
\\
 \nonumber    &&+ (y^2)^2 (\nu -1)
   P_{\nu -2}(z) P_{\nu -1}(z)\Big],
\end{eqnarray}
{which leads to}
\begin{equation}\label{Gamy}
    \Gamma^{11}(z)=-\frac{Q_{\nu }(z)}{P_{\nu }(z)},\
    \Gamma^{12}(z)=\Gamma^{21}(z)=0,\ \Gamma^{22}(z)=-\frac{Q_{\nu
    -2}(z)}{P_{\nu-2}(z)}.
\end{equation}
{In the coordinates $z,w,y^1,y^2$ the dual metric (\ref{dualmetric1}) reads}
\begin{equation}\label{Rosendualmetric}
ds^2= \frac{2}{z^2-1}\, dz\,dw+(P_{\nu}(z))^2(dy^1)^2+(P_{\nu-2}(z))^2(dy^2)^2.
\end{equation}
One can get rid of the denominator of the first term by substitution \\$z= -\tanh\,\tilde
z$
and bring the metric to the diagonal Rosen form. 

Transition to the Brinkman coordinates is obtained \cite{blaulough,blauffpapa}  by virtue
of the matrices $C(\tilde z)$ and $Q(\tilde z)$, where $C(\tilde z)$ is given by the Rosen
form of the matrix; in our case
\begin{equation}\label{matC}
    C(\tilde z)=\left(\begin{array}{cc}
                      ( P_{\nu }(-\tanh\,\tilde z))^2 & 0 \\
                      0 & (P_{\nu-2 }(-\tanh\,\tilde z))^2\end{array}
                   \right),
\end{equation}
and the matrix $Q(\tilde z)$ is a solution of \eqn s \begin{equation}\label{eqnQ}
   Q(\tilde z)\cdot C(\tilde z)\cdot Q(\tilde z)^t=\unit,\ \ \dot Q(\tilde z) \cdot C(\tilde z)\cdot Q(\tilde z)^t = Q(\tilde z)\cdot C(\tilde z)\cdot\dot  Q(\tilde z)^t.
\end{equation}
In our case, the solution can be chosen as
\begin{equation}\label{matQ}
    Q=\left(\begin{array}{cc}
                      ( P_{\nu }(-\tanh\,\tilde z))\-1 & 0 \\
                      0 & (P_{\nu-2 }(-\tanh\,\tilde z))\-1\end{array}
                   \right).
\end{equation}
The Brinkman coordinates {can be written} in terms of the Rosen coordinates as
\begin{eqnarray*}
  x^+ &=& \tilde z ,\\ \nonumber
 x^- &=& w -\frac{1}{2} (\nu +1) P_{\nu }(-\tanh\,\tilde z) \Big
 [P_{\nu +1}(-\tanh\,\tilde z)+ \tanh\,\tilde z P_{\nu }(-\tanh\,\tilde z) \Big](y^1)^2\\
 && -\frac{1}{2} (\nu -1) P_{\nu -2}(-\tanh\,\tilde z) \Big[P_{\nu -1}(-\tanh
   \,\tilde z)+\tanh\,\tilde z P_{\nu -2}(-\tanh\,\tilde z)\Big](y^2)^2 ,\\
  z^1 &=& \,y^1  P_{\nu }(-\tanh\,\tilde z),\\
  z^2 &=&\,y^2 P_{\nu-2 }(-\tanh\,\tilde z).
\end{eqnarray*}
Using moreover the inverse of the transformation of coordinates (\ref{XiInRosen}) we get
\begin{eqnarray}
\nonumber  x^+ &=& -{\rm arctanh}(\tilde x_1) ,\\
\nonumber  x^- &=& - \tilde x_4+\frac{1}{2} \tilde x_1 \left(\tilde x_3^2 (\nu -1)-\tilde
x_2^2 \nu
   \right),\\
\label{BrinkInXi} z^1 &=& \tilde x_2 ,\\
\nonumber z^2 &=& \tilde x_3.
\end{eqnarray}  The dual metric (\ref{dualmetric1}) is then transformed to the form of the
plane--parallel wave (\ref{homogeneousmetric}), where
\begin{equation}\label{aij}
    K_{ij}(x^+)=\frac{1}{(\cosh x^+)^2}\left(\begin{array}{cc}
                       \nu (\nu+1) & 0 \\
                      0 & (\nu-2) (\nu-1)\end{array}
                   \right).
\end{equation}

It means that the non-abelian T-duality transforms the plane--parallel wave metric
(\ref{FF})  to another plane--parallel wave metric that is again solvable. The solutions of
classical equations of motion in the dual background can be obtained by the Ansatz
\begin{equation}\label{dualansatz}
    X^+(\tau,\sigma)=-{\rm arctanh}(\frac{c_1-\kappa \sigma}{\kappa\tau})
\end{equation} that follows from the duality and coordinate transformation of the
light-cone gauge {of} the original background. From the transformation (\ref{BrinkInXi})
and (\ref{X2},\ref{X3}) one can see that the transversal components
$Z^1(\tau,\sigma),Z^2(\tau,\sigma)$ of the classical solutions  of the dual model are again
expressed in terms of the Hankel functions.

The conformal invariance conditions (\ref{bt1})--(\ref{bt3}) written in the Brinkman
coordinates result in very simple \eqn{} for the dilaton
\begin{equation}\label{betaeqnpp3}
  \tilde{\Phi}''(x^+)=\frac{2 (1-\nu+\nu^2)}{\cosh^2(x^+)},
\end{equation}
solved in agreement with (\ref{dualdilatonhomogeneous3}) by
\begin{equation}\label{dilaton3}
     \tilde{\Phi}(x^+)=C_0+C_1\,x^+ + 2(1-\nu+\nu^2)\log(\cosh x^+).
\end{equation}
\section{Strings in the dual background obtained from $\mathfrak{g}_2$}
{Let us now consider the group generated by the Lie algebra
$\mathfrak{g}_2=Span[T_1, T_2, T_3, T_4]$ with commutation relations (cf.
(\ref{cr1}))
\begin{align}
 \nonumber       [T_4,T_{1}]&=T_{1},\\
        [T_4,T_{2}]&=\nu\, T_{2}-\rho\ T_3,\\ \nonumber
        [T_4,T_{3}]&=\nu\, T_{3}+\rho\ T_2.
\end{align}}
The transformation between group coordinates
$x^{1},x^{2},x^{3},x^{4}$ and geometrical coordinates $u,v,x,y$ is
     \begin{align} \label{tfncoorshomogeneous}
\nonumber        u&=e^{x^{4}},\\
        v&=[-\frac{1}{2}\nu((x^{2})^{2}+(x^{3})^{2})+x^{1}]e^{-x^{4}},\\\nonumber
        x&=x^{2}\cos(\rho\, x^4)-x^3\sin(\rho\, x^4),\\\nonumber
        y&=x^{3}\cos(\rho\, x^4)+x^2\sin(\rho\, x^4).
    \end{align}
The metric $(\ref{FF})$ is then transformed into the form
    \be\label{F}
        G_{\mu\nu}(x)=\left(
                     \begin{array}{cccc}
                       0 & 0 & 0 & 1\\
                      0 & 1 & 0 & -\nu x^{2}-\rho\, x^3\\
                       0 & 0 & 1 & -\nu x^{3}+\rho\, x^2 \\
                       1 & -\nu x^{2}-\rho\, x^3 & -\nu x^{3}+\rho\, x^2 & -2x^{1}+(\nu^{2}+\rho^2)((x^{2})^{2}+(x^{3})^{2}) \\
                     \end{array}
                   \right).
    \ee
The equations (\ref{btp},\ref{btm}) for $\tilde h$ now read
\begin{equation}\label{atau2}
    \partial_\tau\tilde h=-\left(
\begin{array}{c}
 0 \\
 (\kappa  \tau )^{\nu }\partial_{\sigma} X \\
 (\kappa  \tau )^{\nu } \partial_{\sigma} Y \\
 \kappa  \tau\,  \partial_{\sigma} V- \rho\, Y\partial_{\sigma} X+  \rho\, X\partial_{\sigma}
 Y
\end{array}
\right),
\end{equation}
\begin{equation}\label{asigma2}
   \partial_\sigma\tilde h= -\left(
\begin{array}{c}
 \kappa  \\
 (\kappa  \tau )^{\nu } \left(\partial_{\tau} X-\frac{\nu}{\tau}  X\right)\\
 (\kappa  \tau )^{\nu } \left(\tau  \partial_{\tau} Y-\frac{\nu}{\tau} Y\right) \\
 \kappa  \tau \partial_{\tau} V-\kappa   V-\frac{\nu(1-\nu)}{\tau }( X^2+ Y^2)
 - \rho\, Y\partial_{\tau} X+  \rho\, X\partial_{\tau}
 Y
\end{array}
\right),
\end{equation}
and are solved by
\begin{eqnarray}
\label{lcsolnRho}  \tilde h_1 &=&c_1-\kappa\sigma  ,\\
\label{h2Rho}  \tilde h_2 &=& c_2 -(\kappa\tau)^\nu\int d\sigma(\partial_\tau X-\frac{\nu}{\tau}X),\\
\label{h3Rho}  \tilde h_3 &=& c_3 -(\kappa\tau)^\nu\int d\sigma(\partial_\tau Y-\frac{\nu}{\tau}Y),\\
  \tilde h_4 &=& c_4 +\int d\sigma\Big[\frac{\nu(1-\nu)}{\tau}(X^2+Y^2)+\nonumber \\ && \kappa(V-\tau\partial_\tau
  V) +  \rho\, Y\partial_{\tau} X-  \rho\, X\partial_{\tau}
 Y\Big],
\end{eqnarray}
{where $c_i, i=1,\ldots,4$ are arbitrary constants.}

The tensor dual to (\ref{F}) is \cite{hlatur3}
\begin{equation}\label{dualhomogenoeous} \tilde
F_{\mu\nu}(\tilde{x})= \left(
\begin{array}{cccc}
 \frac{(\nu ^2+\rho^2) \left(\tilde{x}_2^2+\tilde{x}_3^2\right)}{\tilde{x}_1^2-1} & \frac{\nu\,  \tilde{x}_2-\rho\,\tilde{x}_3}{1-\tilde{x}_1} &
 \frac{\nu\,
   \tilde{x}_3+\rho\,\tilde{x}_2}{1-\tilde{x}_1} & \frac{1}{1-\tilde{x}_1} \\
 \frac{-\nu\,  \tilde{x}_2+\rho\,\tilde{x}_3}{\tilde{x}_1+1} & 1 & 0 & 0 \\
 \frac{-\nu\,  \tilde{x}_3-\rho\,\tilde{x}_2}{\tilde{x}_1+1} & 0 & 1 & 0 \\
 \frac{1}{\tilde{x}_1+1} & 0 & 0 & 0
\end{array}
\right).
\end{equation}
It has nontrivial antisymmetric part (torsion potential)  \be
 \tilde{B}_{\mu\nu}=\frac{1}{2}(\tilde F_{\mu\nu}-\tilde F_{\nu\mu})\,
\label{torpot} \ee with torsion $\tilde{H}=d\tilde{B}$ {equal to}
\begin{equation}\label{dualtorsionhom}
   \tilde{H}= \frac{2\rho}{\tilde{x}_1^2-1}\,d\tilde{x}_1\wedge d\tilde{x}_2\wedge d\tilde{x}_3.
\end{equation}
The dual metric, which is the symmetric part of (\ref{dualhomogenoeous}), does not solve the
Einstein \eqn s but satisfies the conformal invariance conditions (\ref{bt1})--(\ref{bt3}) with
the dilaton field
\begin{equation}\label{dualdilatonhomogeneous2} \tilde{\Phi}= C_0 +
C_1 \ln\left(\frac{\tilde{x}_1-1}{\tilde{x}_1+1}\right) -\nu(\nu+1)\ln(1-\tilde{x}_1^2).
\end{equation}
The equations (\ref{lghtgth}) are solved {using the same faithful representation of the
Drinfeld double} as in the previous section. From (\ref{replgh}) we get {the relation
between the solutions of the dual sigma models as}
\begin{eqnarray}
h^j&=&x^j \label{tq0} ,\\
\tilde{x}_1&=& e^{-x^4} {\tilde h}_1,\label{tq1}\\
\tilde{x}_2&=&e^{-\nu x^4} ({\tilde h}_3 \sin (\rho x^4)+{\tilde h}_2 \cos (\rho x^4)),\label{tq2}\\
\tilde{x}_3&=&e^{-\nu x^4} ({\tilde h}_3 \cos (\rho x^4)-{\tilde h}_2 \sin (\rho x^4)),\label{tq3}\\
\tilde{x}_4&=& \nonumber e^{-\nu x^4} (\nu x^3-\rho x^2) ({\tilde h}_3 \cos
   (\rho x^4)-{\tilde h}_2 \sin (\rho x^4))\\ &&+e^{-\nu x^4} (\nu x^2+\rho x^3) ({\tilde h}_3 \sin (\rho x^4)+{\tilde h}_2 \cos (\rho x^4)) \\
    \nonumber &&+e^{-x^4} x^1 {\tilde h}_1+{\tilde h}_4.
\end{eqnarray}
The equation of motion $$ \frac{\delta S_{\wt F}}{\delta \wt X_4} =0 $$ has the same form
(\ref{eqnfortx1}) as in the previous section, and from (\ref{lcsolnRho}, \ref{tq1}) we get again its
solution\begin{equation}\label{stq1} \tilde X_1(\tau,\sigma) = \frac{c_1-\kappa\sigma
}{\kappa \tau}.
\end{equation}The other two \eqn s then reduce to
\begin{eqnarray}\label{eqnfortq2}
(\partial_\sigma^2-\partial_\tau^2)\wt X_2 +\frac{2\rho}{\tau}\partial_\tau\wt X_3+
 \frac{1}{\tau^2}\left[(\nu+\nu^2+\rho^2)\wt X_2-\rho\wt X_3\right]= 0,\\ \label{eqnfortq3}
(\partial_\sigma^2-\partial_\tau^2)\wt X_3 -\frac{2\rho}{\tau}\partial_\tau\wt X_2+
 \frac{1}{\tau^2}\left[(\nu+\nu^2+\rho^2)\wt X_3+\rho\wt X_2\right]= 0.\end{eqnarray}
Their solution for $\nu\neq \half$ follows from (\ref{tq2}), (\ref{tq3}) and \eqref{h2Rho},
\eqref{h3Rho} in the form  \begin{eqnarray}\nonumber
  \tilde X_2(\tau,\sigma) &=&\cos\Big(\rho\,\log(\kappa \tau)\Big)
  \Big[ ( {c_2}\, \kappa^{-\nu }+\sqrt{2 \nu -1}\, \tilde x\,\sigma)\tau^{-\nu}+
  \Sigma_2^x(\tau,\sigma) \Big]\ +\\ &&
    \sin\Big(\rho\,\log(\kappa \tau)\Big) \Big[ ( {c_3}\, \kappa^{-\nu }-\sqrt{2 \nu -1}\
    \tilde y\,\sigma)\tau^{-\nu}+
  \Sigma_2^y(\tau,\sigma) \Big],\label{solx2rho}\\\nonumber
\tilde X_3(\tau,\sigma) &=&
  -\sin\Big(\rho\,\log(\kappa \tau)\Big)\Big[ ( {c_2}\, \kappa^{-\nu }+\sqrt{2 \nu -1}\, \tilde x\,\sigma)\tau^{-\nu}+
  \Sigma_2^x(\tau,\sigma) \Big]\ +\\ &&
    \cos\Big(\rho\,\log(\kappa \tau)\Big) \Big[ ( {c_3}\, \kappa^{-\nu }-\sqrt{2 \nu -1}\
    \tilde y\,\sigma)\tau^{-\nu}+
  \Sigma_2^y(\tau,\sigma) \Big],\label{solx3rho}\end{eqnarray}where $\Sigma_2^j$  {are} given by (\ref{Sigma2}).
{The last equation of motion is solved by}
 \be\nonumber
\tilde X_4(\tau,\sigma) = c_4-\frac{\sigma + c_1}{ \tau}x^1+\  (\nu x^2+ \rho x^3)\ \tilde X_2+(\nu x^3 - \rho x^2)\ \tilde X_3 \ee\be + \int d\sigma[\frac{\nu(1-\nu)}{\tau}(X^2+Y^2)+\kappa(V-\tau\partial_\tau
  V) + \rho Y \partial_{\tau}X-\rho X \partial_{\tau}Y].\ee

{It is worth mentioning that the duality is not the only option here. It is possible to use
another decomposition \cite{sno:priv} of $D$ into groups $\hat{G}, \bar{G}$ with algebra
$\mathfrak d=\{\hat e_1,\ldots,\hat e_4,\bar e_1,\ldots,\bar e_4\}=\{T_1+T_4,\tilde
T^1-\tilde T^4,T_2,T_3,\half (\tilde T^1+\tilde T^4),\half ( T_1- T_4),\tilde T^2, \tilde
T^3\}$ and commutation relations of the basis elements
\begin{align*}
[\hat e_1, \hat e_2]&=-\hat e_2, & [\bar e_1,\bar e_2]&=-\half \bar e_1, \\
[\hat e_1, \hat e_3]&=\nu \hat e_3 - \rho \hat e_4, & [\bar e_2,\bar e_3]&=\half \nu \bar e_3 + \half \rho \bar e_4,\\
[\hat e_1, \hat e_4]&=\rho \hat e_3 + \nu \hat e_4, & [\bar e_2,\bar e_4]&=-\half \rho \bar
e_3 + \half \nu \bar e_4.
\end{align*}
Then it is possible to construct sigma models on the groups $\hat G$ or $\bar G$ and in principle
solve their equations of motion by the Poisson--Lie T-plurality transformation
\cite{unge}, which relates the solutions of the sigma models on $G$ and $\hat G$ or $\bar G$ respectively. However,
all the calculations get very complicated as none of the algebras is Abelian.}

\subsection{Killing vectors for the dual metric and its
 plane--parallel wave form} In general there are only two linearly
independent Killing vectors of the tensor (\ref{dualhomogenoeous}) satisfying ${\mathcal
L}_{\tilde K}\tilde F=0$, namely
\begin{eqnarray}
 \tilde K_1 &=& \tilde x_2\frac{\partial}{\partial \tilde x_3}
  -\tilde x_3\frac{\partial}{\partial \tilde x_2},\\
  \tilde K_2 &=& -\frac{\partial}{\partial \tilde x_4}.
\end{eqnarray}

However, for $\rho=0$ the torsion (\ref{dualtorsionhom}) vanishes and both the \eqn s of
motion and the $\beta$ \eqn s (\ref{bt1})--(\ref{bt3}) for the \sm {} are equivalent to
those calculated from the symmetric part of (\ref{dualhomogenoeous})
\begin{equation}\label{dualmetric2}
\tilde  G_{\mu\nu}(\tilde{x})=   \left(
\begin{array}{cccc}
 \frac{\left({\tilde  x_2}{}^2+{\tilde  x_3}{}^2\right) \nu ^2}{{\tilde  x_1}{}^2-1}
 & \frac{\nu\,  {\tilde  x_1}{} {\tilde  x_2}{}}{1-{\tilde  x_1}{}^2} &
   \frac{\nu\, {\tilde  x_1}{} {\tilde  x_3}{}}{1-{\tilde  x_1}{}^2} & \frac{1}{1-{\tilde  x_1}{}^2} \\
 \frac{\nu\,  {\tilde  x_1}{} {\tilde  x_2}{}}{1-{\tilde  x_1}{}^2} & 1 & 0 & 0 \\
 \frac{\nu\, {\tilde  x_1}{} {\tilde  x_3}{}}{1-{\tilde  x_1}{}^2} & 0 & 1 & 0 \\
 \frac{1}{1-{\tilde  x_1}{}^2} & 0 & 0 & 0
\end{array}
\right).
\end{equation}

We shall show that this metric is again a  plane--parallel wave. It has a six-dimensional
algebra of Killing vectors generated by $ \tilde K_1,\tilde K_2 $ and
\begin{eqnarray}
\nonumber  \tilde K_3&=& P_\nu(\tilde  x_1)\frac{\partial}{\partial \tilde x_2}
  +\tilde x_2[(\nu+1)P_{\nu+1}(\tilde  x_1)-\tilde  x_1(1+2\nu)P_\nu(\tilde  x_1)]\frac{\partial}{\partial \tilde x_4} ,\\
\nonumber  \tilde K_4&=&  Q_\nu(\tilde  x_1)\frac{\partial}{\partial \tilde x_2}
  +\tilde  x_2[(\nu+1)Q_{\nu+1}(\tilde  x_1)-\tilde  x_1(1+2\nu)Q_\nu(\tilde  x_1)]\frac{\partial}{\partial \tilde x_4} ,\\
 \tilde K_5&=& P_{\nu}(\tilde  x_1)\frac{\partial}{\partial \tilde x_3}
  +\tilde  x_3[(\nu+1)P_{\nu+1}(\tilde  x_1)-\tilde  x_1(1+2\nu)P_\nu(\tilde  x_1)]\frac{\partial}{\partial \tilde x_4} ,\\
\nonumber   \tilde K_6&=& Q_{\nu}(\tilde  x_1)\frac{\partial}{\partial \tilde x_3}
  +\tilde  x_3[(\nu+1)Q_{\nu+1}(\tilde  x_1)-\tilde  x_1(1+2\nu)Q_\nu(\tilde  x_1)]\frac{\partial}{\partial \tilde x_4}
.\end{eqnarray} Their nonzero commutation relations are
\begin{align}
\nonumber [\tilde K_1,\tilde K_3]&=-\tilde K_5 &
\nonumber [\tilde K_1,\tilde K_4]&=-\tilde K_6,\\
\label{ivocr} [\tilde K_1,\tilde K_5]&=\tilde K_3, & [\tilde K_1,\tilde K_6]&=\tilde K_4 ,
\\ \nonumber [\tilde K_3,\tilde K_4]&=\tilde K_2, & [\tilde K_5,\tilde K_6]&=\tilde K_2.
\end{align}
One can see that the Killing vectors  $\tilde K_2$--$\tilde K_6$ form the Heisenberg
algebra {with the central element $\tilde K_2$. This opens the possibility that this metric
might be again} brought to the form of a  plane--parallel wave. The transformation to Rosen
coordinates $z,w,y^1,y^2$
\begin{eqnarray}
 \nonumber  \tilde x_1 &=& z ,\\
 \nonumber  \tilde x_2  &=& y^1 P_\nu(z),\\
 \label{XiInRosenRho}  \tilde x_3  &=& y^2 P_{\nu}(z),\\
 \nonumber \tilde x_4 &= & -w + \half\Big[(y^1)^2+(y^2)^2\Big] P_{\nu }(z) \Big[(\nu +1) P_{\nu +1}(z)-z(1+ 2 \nu) P_{\nu
 }(z)\Big]
\end{eqnarray}
brings the {Killing vectors $\tilde K_2$--$\tilde K_6$ to} the form (\ref{kilbffpp}) and
the metric to
\begin{equation}\label{Rosendualmetric2}
ds^2= \frac{2}{z^2-1}\, dz\,dw+(P_{\nu}(z))^2(dy^1)^2+(P_{\nu}(z))^2(dy^2)^2.
\end{equation}

Transition to the Brinkman coordinates is obtained similarly as in the section \ref{42}
\begin{eqnarray}
\nonumber  x^+ &=& -{\rm arctanh}(\tilde x_1) ,\\
\nonumber  x^- &=& - \tilde x_4-\frac{1}{2}\,\nu\, \tilde x_1 (\tilde x_2^2 +\tilde x_3^2),
\\
\label{BrinkInXi2} z^1 &=& \tilde x_2 ,\\
\nonumber z^2 &=& \tilde x_3,
\end{eqnarray}  and the dual metric (\ref{dualmetric2}) is transformed to the form of the
plane--parallel wave \be \label{homogeneousmetricr0}
        ds^{2}=2dx^+dx^- - \frac{ \nu (\nu+1)[(z^1)^2+(z^2)^2]}{(\cosh
        x^+)^2}(dx^+)^2+(dz^1)^2+(dz^2)^2
    \ee
that is isotropic in $z^1,z^2$.

Similarly for  $\rho\neq 0$ we can use a rotated version of (\ref{BrinkInXi2})
\begin{eqnarray}
\nonumber  x^+ &=& -{\rm arctanh}(\tilde x_1) ,\\
\nonumber  x^- &=& - \tilde x_4-\frac{1}{2}\,\nu\, \tilde x_1 (\tilde x_2^2 +\tilde x_3^2),
\\
\label{BrinkInXi3} z^1 &=& \tilde x_2\cos\Omega-\tilde x_3\sin\Omega ,\\
\nonumber z^2 &=& \tilde x_2\sin\Omega+\tilde x_3\cos\Omega,
\end{eqnarray} where $\Omega= \rho\,\log(\cosh\ x^+)$, to bring  the dual metric {derived from (\ref{dualhomogenoeous})} to the form of the
plane--parallel wave in Brinkman coordinates  \begin{eqnarray}
 \label{homogeneousmetricr3}
\nonumber        ds^{2}&=& -[(z^1)^2+(z^2)^2] \frac{ 2 \nu (\nu+1)+\rho^2(1+\cosh(2x^+))}{2(\cosh
        x^+)^2}(dx^+)^2\\&&+ 2dx^+dx^-+(dz^1)^2+(dz^2)^2.
    \end{eqnarray}
The torsion becomes constant in these coordinates
$$ \tilde{H}= 2\rho dx^+\wedge dz^1\wedge dz^2,$$
and the vanishig $\beta $ equations (\ref{bt1})--(\ref{bt3}) result in the \eqn{} for the
dilaton
\begin{equation}\label{betaeqnpp3}
  \tilde{\Phi}''(x^+)=\frac{2 \nu (\nu+1)}{\cosh^2(x^+)},
\end{equation}
solved in agreement with (\ref{dualdilatonhomogeneous2}) by
\begin{equation}\label{dilaton3}
     \tilde{\Phi}(x^+)=C_0+C_1\,x^+ + 2 \nu (\nu+1)\log(\cosh x^+).
\end{equation}

 {Thus we can see that the non-abelian T-duality based on the
semi-abelian \dd {} given by (\ref{cr1}) transforms} the plane--parallel wave metric
(\ref{FF}) to another plane--parallel wave metric \eqref{homogeneousmetricr3} and torsion.
\section{Conclusions}{\tuc We have investigated non--Abelian T--duals  of homogeneous pp-wave
metric \eqref{FF} that belongs to the class of string backgrounds with metric, B--field and
dilaton  \begin{eqnarray}
  ds^{2}&=& 2dudv-K_{ij}(u)x^{i}x^{j}du^{2}+d\vec{x}^{2}, \label{concl1} \\
  B&=&\half H_{ij}(u)x^i du\wedge dx^j,\ \ \Phi=\Phi(u)\label{concl2}\end{eqnarray}
 analyzed in \cite{BLPT}.
We have found classical string solutions for the dual backgrounds and we have also shown
that by appropriate coordinate transformations the dual backgrounds can be brought again to
the form \eqref{concl1}, \eqref{concl2}. The number of Killing vectors of the dual models
is less than those of (\ref{FF}), which means {that the dual backgrounds are different}
from the initial one.

The dual metrics  (\ref{dualmetric1}) and (\ref{dualmetric2}),  obtained in the group
coordinates by the procedure described in the Section \ref {secPLT}, have five-dimensional
subalgebras of Killing vectors commuting as the two-dimensional Heisenberg algebra. This
implies {that by a change of coordinates they} can be rewritten to the diagonal Rosen form
or plane--parallel wave form given by (\ref{aij}) and (\ref{homogeneousmetricr0}). By
virtue of more general coordinate \tfn{} \eqref{BrinkInXi3} we find that the metric
obtained from the dual tensor \eqref{dualhomogenoeous} can be brought to the plane-parallel
wave form
\eqref{homogeneousmetricr3} even for $\rho\neq 0$. 
The conformal invariance conditions (\ref{bt1})--(\ref{bt3}) for the dual backgrounds in
Brinkman coordinates then acquire simple form of solvable ODE \cite{BLPT} for the dilaton
\begin{equation}\label{blpt}
    \Phi''(u)=K_{ii}(u)+H_{ij}(u)H_{ij}(u),
\end{equation}
and the dual backgrounds satisfy the conformal invariance as well.}

The classical string solutions of the dual models, similarly as in the case of the initial
metric (\ref{FF}), are given in terms of Hankel functions. An interesting point is that the
equations of dual models may not admit the usual light-cone gauge, but the light-cone gauge
(\ref{lightcg}) is transformed to the solution (\ref{tx1u}) of the \eqn{}
(\ref{eqnfortx1}). On the other hand, notice that the backgrounds \eqref{aij} and
\eqref{homogeneousmetricr3} obtained by coordinate transformations of the dual backgrounds
do admit the light-cone gauge. However, in these cases it leads to complicated equations
for the transversal components. In order to get the solution, the use of Ansatz
(\ref{dualansatz}) is crucial.

\section*{Acknowledgement}
This work was supported by the Grant Agency of the Czech Technical University in Prague,
grant No. SGS13/217/OHK4/3T/14 and by the research plan MSM6840770039 of the Ministry of
Education of the Czech Republic.

\end{document}